\documentclass[twocolumn,aps,pra,superscriptaddress]{revtex4-1}

\usepackage[utf8x]{inputenc}
\usepackage{amsmath}
\usepackage{amssymb}
\usepackage{graphicx}
\usepackage{upgreek}
\usepackage{bm}
\usepackage{float}
\usepackage{xcolor}
\usepackage{soul}

\newcommand{\vect}[1]{\bm{#1}}
\newcommand{\ten}[1]{\mbox{\textbf{{\textsf{#1}}}}}

\newcommand{\sprod}{\!\cdot\!}

\newcommand{\dif}{\mathrm{d}}
\newcommand{\mi}{\mathrm{i}}

\newcommand{\new}[1]{\textcolor{black}{#1}}

\begin{document}
\title{Casimir--Polder interaction of neutrons with metal or
dielectric surfaces}

\author{Valentin Gebhart}
\altaffiliation[Current address: ]{QSTAR, INO-CNR and LENS, Largo Enrico Fermi 2, 50125 Firenze, Italy}
\affiliation{Physikalisches Institut, Albert-Ludwigs-Universit\"at
Freiburg, Hermann-Herder-Str.~3, 79104 Freiburg, Germany}

\author{Juliane Klatt} 
\altaffiliation[Current address: ]{Department for Biosystems Science and Engineering, ETH Z\"urich, Mattenstr.
26, 4058 Basel, Switzerland}
\affiliation{Physikalisches Institut, Albert-Ludwigs-Universit\"at
Freiburg, Hermann-Herder-Str.~3, 79104 Freiburg, Germany}

\author{Gunther Cronenberg}
\affiliation{Atominstitut, Technische Universit\"at Wien, Stadionallee~2, 1020 Wien, Austria}

\author{Hanno Filter}
\affiliation{Atominstitut, Technische Universit\"at Wien, Stadionallee~2, 1020 Wien, Austria}
\affiliation{Technische Universit\"at M\"unchen, Physikdepartment - E66, Boltzmannstr. 2, 85748 Garching, Germany}

\author{Stefan Yoshi Buhmann}
\email{stefan.buhmann@uni-kassel.de}
\affiliation{Physikalisches Institut, Albert-Ludwigs-Universit\"at
Freiburg, Hermann-Herder-Str.~3, 79104 Freiburg, Germany}
\affiliation{Institut für Physik, Universit\"at Kassel, Heinirich-Plett-Str. 40, 34132 Kassel, Germany}

\date{\today}

\begin{abstract}
We predict a repulsive Casimir--Polder-type dispersion interaction
between a single neutron and a metal or dielectric surface. We consider a
scenario where a single neutron is subject to an external magnetic
field. Due to its intrinsic magnetic moment, the neutron then forms
a magnetisable two-level system which can exchange virtual photons
with a nearby surface. The resulting dispersion interaction between a
purely magnetic object (neutron) and a purely electric one (surface)
is found to be repulsive, in contrast to the typical attractive interaction between electric objects. Its magnitude is considerably smaller than
the standard atom--surface Casimir--Polder force due to the magnetic 
nature of the interaction and the smallness of the electron-to-neutron 
mass ratio. Nevertheless, we show that it can be comparable to the 
gravitational potential of the same surface and should be taken 
into consideration in future neutron interference experiments.
\end{abstract}
%\pacs{
%12.20.--m, % Quantum electrodynamics
%42.50.Ct,  % Quantum description of interaction of light and matter;
           % related experiments
%42.50.Nn,  % Quantum optical phenomena in absorbing, amplifying,
           % dispersive and conducting media; cooperative phenomena in
           % quantum optical systems
%03.75.Dg   % Atom and neutron interferometry
%}

\maketitle

%\tableofcontents

\section{Introduction}

As originally conceived by Casimir, the attractive electromagnetic force between two
perfectly conducting parallel plates is a consequence of the quantum
fluctuations of the electromagnetic field which persist even when the
field is in its vacuum state of zero temperature \cite{0373}. The
plates, which are merely loci of boundary conditions supporting
standing-wave modes of the electromagnetic field in Casimir's 
picture, are assigned a much more active role in Lifshitz' theory for 
two dielectric plates \cite{0057}: here, the fluctuating polarisation
within the dielectric media ultimately generates the force. It is 
hence
apparent that dispersion forces may much more generally arise as
effective electromagnetic forces between any polarisable objects. They
may be attributed to quantum zero-point fluctuations of the objects'
polarisation and of the electromagnetic field \cite{0487}. In
particular, the term Casimir--Polder force is commonly used to refer
to the dispersion interaction between a microscopic object such as an
atom or a molecule and a macroscopic body \cite{0030}.

Shortly after Casimir's seminal work, it was predicted by Boyer that 
the force between a perfectly conducting plate and an infinitely 
permeable one is repulsive \cite{0122}. Mathematically, this is due to 
the different boundary conditions that electric vs magnetic mirrors 
place on the electromagnetic field \cite{0123}: the force depends on 
the product of the reflection coefficients of the two plates and is 
hence attractive for two electric or two magnetic mirrors and 
repulsive for two mirrors of different type. Repulsive dispersion 
forces have since been predicted for a variety of scenarios 
involving a polarisable and a magnetisable object 
\cite{0089,0838,0134,0126}, including the Casimir--Polder force 
between an atom and a plate
\cite{0095,0330,Henkel05,0019,0831,Haakh09,Bimonte09}. While the 
attractive Casimir--Polder force between a polarisable atom and a 
perfect electric mirror is a straightforward consequence of the 
attractive alignment of the fluctuating atomic dipole moment and its 
image \cite{0022}, an understanding of the repulsion for mixed 
electric--magnetic object combinations requires electrodynamical 
considerations. As explicitly shown for the case of two atoms, an 
oscillating electric dipole generates a magnetic field which orients a 
nearby magnetic dipole such that a repulsive force emerges 
\cite{0096,0121}.

The study of repulsive dispersion forces is motivated by the hope
that these could help overcome the problem of stiction in
nanotechnology \cite{0578}. Theoretical studies have unearthed three
mechanisms by which repulsion can be achieved, two of which have been
verified experimentally: (i) two bodies immersed in a liquid repel
each other when one of them is more optically thin and the other more
optically thick than the medium \cite{0961}, the effect being
analogous to an air bubble in water experiencing `repulsive gravity'.
(ii) Non-equilibrium systems such as non-uniform temperatures
\cite{0771} or excited atoms in a low-temperature environment
\cite{0157} may experience repulsion, which is analogous to the force
that an oscillating dipole exerts on an second, out-of-phase dipole of
lower eigenfrequency. (iii) The mentioned repulsion due to magnetic
properties has proven elusive so far, because for materials existing
in nature it is typically overwhelmed by the ever-present attractive
electric--electric force \cite{0134}. Attempts to overcome this
problem via artificial meta-materials \cite{0126,0953,0954} have
been demonstrated to fail due to an Earnshaw no-go theorem
\cite{0946}.

Here, we propose a system that is free from such constraints, because
one of the interacting partners---a neutron---is purely magnetic.
While electrically neutral and non-polar, the neutron does exhibit a
magnetic moment which may interact with the quantum electrodynamic
field. As we will argue, the neutron with its spin eigenstates can be
viewed as a magnetisable two-level system which will experience a
repulsive force of Casimir--Polder type when interacting with a 
metal or dielectric wall. To our knowledge, this setup is the only physical configuration of magnetic (neutron) and electric (surface) objects that leads to an instrinsic repulsive interaction, since in all other cases the attractive electric--electric interaction is dominating. Note that for the case of a perfectly 
conducting wall, such an interaction of a spin particle has been 
studied extensively by Babiker and Barton 
\cite{Babiker72}. Van der Waals-type neutron--neutron 
interactions have been investigated very recently by Babb and Hussein 
\cite{Babb16}.

In view of practical applications, neutrons as probes allow for very clean, systematic experiments due to their limited ability to couple to the environment.
They carry neither a measurable nonzero electrical charge nor an 
electric dipole moment.
The current upper limit for an electric charge $q_n$ is given 
by $|q_n|<1.3\cdot10^{-21}\, q_e$~\cite{Baumann:1988} ($q_e$ is the electric charge), while the current upper limit for the magnitude electric dipole-moment $d_n$ is 
$|d_n|<1.8\cdot10^{-26}\, q_e\, \mathrm{cm}$~\cite{Abel2020}.
Effects due to a magnetic coupling of the spin of the neutron can 
be effectively shielded, see, e.g., reference~\cite{Lins:2015} for state-of-the-art shielding in a neutron electric dipole moment experiment. Finally, 
its relative long-life time (over $14\,$min) does not constrain 
experimental practicability.

Evidence for contact-interactions of neutrons with surfaces has been found
within the context of neutron interferometry. In particular, by
introducing a stack of narrow slits into one arm of such an
interferometer, a confinement-induced phase shift has been found
\cite{Rauch02}. Here, the plates forming the slits provide rigid
boundary conditions for the neutron wave function. This is in contrast
to our proposed long-range Casimir--Polder interaction which should be
felt by the entire neutron wave function within such slits.

Our analysis yields an important answer to the question whether the Casimir--Polder interaction has to be taken into account in different neutron-based experimental setups: while for current experiments the obtained Casimir--Polder energies lie below the detection threshold, a noticable contribution to future highly-sensitive experiments is conceivable.

The article is organised a follows: we begin by describing the
proposed setup of a neutron in front of a metal or dielectric surface
and introduce the basic formalism of macroscopic quantum
electrodynamics used to describe the surface-assisted magnetic field.
We then derive the Casimir--Polder force on the neutron using
second-order perturbation theory. Finally, we quantify the resulting
neutron--plate interaction for different substrate materials,
compare it to gravitational potentials \new{as well as contributions due to the neutron's static polarisabilities} and discuss its relevance in state-of-the-art gravitational resonance spectroscopy experiments.

%%%%%%%%%%%%%%%%%%%%%%%%%%%%%%%%%%%

\section{Setup and basic equations}
\begin{figure}
\centering
\includegraphics[width=0.45\textwidth]{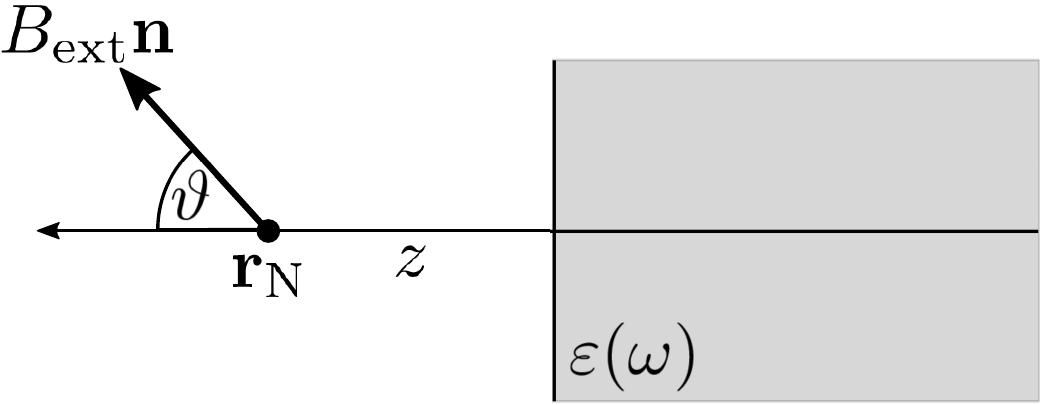}
\caption{Setup: Neutron in front of an infinite metal or dielectric
plate. To lift the degeneracy of the two neutron spin states, an
external magnetic field is applied in a direction
$\vect{n}(\vartheta)$ which is at an angle $\vartheta$ with respect
to the surface normal.}
\label{fig:0}
\end{figure}

As illustrated in Fig.~\ref{fig:0}, we consider a single neutron at
position $\vect{r}_\mathrm{N}$, which is at a distance $z$ from a
homogeneous metal or dielectric plate of electric permittivity
$\varepsilon(\omega)$ which is infinitely thick and infinitely
extended in the lateral dimensions (semi-infinite half space).
\new{The neutron's spin couples to the plate-assisted quantum electromagnetic field that we will assume to be in its ground state. In addition, we consider the presence of a (homogeneous and static) external magnetic field $\vect{B}_\mathrm{ext}$. Being homogeneous in space, $\vect{B}_\mathrm{ext}$ induces no direct force on the neutron. Rather, it serves as an experimentally tunable external control parameter.}

In order to obtain the Casimir--Polder potential of the neutron, we
separate the \new{total} Hamiltonian into the \new{individual} Hamiltonians of the neutron \new{(in the external field $\vect{B}_\mathrm{ext}$)} and
the medium-assisted electromagnetic field on one hand, and the
interaction Hamiltonian on the other. We treat the latter as a
perturbation. The free field Hamiltonian is given by \cite{Buhmann12}
\begin{equation}
\hat{H}_\mathrm{F}=\int
\mathrm{d}^3\vect{r}\int_{0}^{\infty}\mathrm{d}\omega \hbar \omega
\hat{\vect{f}}^\dagger(\vect{r},\omega)\cdot
\hat{\vect{f}}(\vect{r},\omega)\,.
\end{equation} 
Here, $\hat{\vect{f}}^\dagger(\vect{r},\omega)$ are
bosonic creation operators of effective medium--field excitations.
The external (classical) static magnetic field
$\vect{B}_\mathrm{ext}=B_\mathrm{ext}\vect{n}$ splits the
energies of the two neutron spin states. Its directional unit vector
$\vect{n}$ is at an angle $\vartheta$ with respect to the unit normal
of the plate. The resulting Hamiltonian of
the \new{neutron in the external magnetic field} reads
\begin{equation}
\hat{H}_\mathrm{N}=E_\uparrow \left| \uparrow\right\rangle\left\langle
\uparrow\right|
+E_\downarrow \left|
\downarrow\right\rangle\left\langle
\downarrow\right|,
\end{equation}
with energies $E_{\uparrow/\downarrow}=\pm(\hbar \gamma_\mathrm{N}
B_\mathrm{ext})/2$. Here, $\gamma_\mathrm{N}$ is the gyromagnetic
ratio of the
neutron, given by $\gamma_\mathrm{N}=(g_\mathrm{N}e)/(2m_\mathrm{N})$
and relates the magnetic
dipole moment $\hat{\vect{m}}$ of the neutron to its spin
$\hat{\vect{s}}$ via $\hat{\vect{m}}=\gamma_\mathrm{N}\hat{\vect{s}}$.
$g_\mathrm{N}$ is
the $g$-factor of the neutron, $m_\mathrm{N}$ its mass and $e$ the
elementary
electric charge.
Finally, the interaction Hamiltonian is given by
\cite{Buhmann12}
\begin{equation}
\label{int}
\hat{H}_\mathrm{int}=-\hat{\vect{m}}\cdot\hat{\vect{B}}
(\vect{r}_\mathrm{N})
\end{equation}
where $\hat{\vect{B}}$ is the quantised (fluctuating) plate-assisted magnetic field
\begin{multline}
 \label{B}
\hat{\vect{B}}(\vect{r})=\sqrt{\frac{\hbar}{\pi\varepsilon_0}}
\int_{0}^{\infty}\mathrm{d}\omega\,\frac{\omega}{c^2}
\int\limits\mathrm{d}^3\vect{r}'
\sqrt{\mathrm{Im}\varepsilon(\vect{r'},\omega)}\\
\cdot\vect{\nabla} \times
\ten{G}(\vect{r},\vect{r}',\omega) \cdot
\hat{\vect{f}}(\vect{r}',\omega)+\mathrm{h.c.}
\end{multline}
Here, $\ten{G}(\vect{r},\vect{r}',\omega)$ is the dyadic Green's
tensor for the classical electromagnetic field \new{that solves the classical boundary problem of the infinite half-space. In other words, the plate represents classical boundary conditions that influence the modes of the quantuised magnetic field $\hat{\vect{B}}$ by means of the classical Green's tensor $\ten{G}(\vect{r},\vect{r}',\omega)$}. The Green's tensor fulfils the
integral relation
\begin{multline}
\label{2.150}
\frac{\omega^2}{c^2}\int\dif^3s\,
\operatorname{Im}\,\varepsilon(\vect{s},\omega)\,
 \ten{G}(\vect{r},\vect{s},\omega)
 \sprod\ten{G}^\ast(\vect{s},\vect{r}',\omega)\\
 =\operatorname{Im}\,\ten{G}(\vect{r},\vect{r}',\omega)\,.
\end{multline}

%%%%%%%%%%%%%%%%%%%%%%%%%%%%%%%%%%%%%%%

\section{Casimir--Polder potential}
Starting from an uncoupled
state $\left| \{0\}\right\rangle\left| i\right\rangle$, where $\left|
\{0\}\right\rangle$ is the vacuum state of the electromagnetic field
and $i\in\{\downarrow,\uparrow\}$, we use second-order
perturbation theory to find its energy shift
\begin{multline}
U_i= \sum_{k=\uparrow,\downarrow}
\mathcal{P}\int_{0}^{\infty}\mathrm{d}\omega\dfrac{1}{
-\hbar(\omega+\omega_{ki})}\\ \cdot
\int\mathrm{d}^3\vect{r}\left| \left\langle i \right|\left\langle
\{0\} \right| -\hat{\vect{m}}\cdot \hat{\vect{B}}(\vect{r}_\mathrm{N})
\left| \vect{1}(\vect{r},\omega) \right\rangle\left| k
\right\rangle \right| ^2
\end{multline}
where we have defined $\omega_{ik}=(E_i-E_k)/\hbar$. We
decompose the potential into $U_i=U_{i\downarrow}+U_{i\uparrow}$,
where the two terms represent the intermediate state being the
spin-down and the spin-up states of the neutron respectively. Using
equations~(\ref{int}) and (\ref{B}) to evaluate the matrix elements of the
interaction Hamiltonian, combining the results by means of the
integral equation (\ref{2.150}) and exploiting Cauchy's integral
formula, one finds
\begin{align}\label{dduu}
&U_{\downarrow\downarrow}=U_{\uparrow\uparrow}\\\nonumber
&\hspace{3.7ex}=\dfrac{\mu_0}{2}\vect{m}_{\downarrow \downarrow} 
\cdot  
\vect{\nabla}\times 
\ten{G}^{(1)}(\vect{r}_\mathrm{N},\vect{r}_\mathrm{N},0) \times 
\overleftarrow{\vect{\nabla'}}\cdot 
\vect{m}_{\downarrow\downarrow}\,,\\\label{du}
&U_{\downarrow\uparrow}=\frac{\mu_0}{\pi}\int_{0}^{\infty}\mathrm{d}
\xi\dfrac{\omega_{\uparrow\downarrow}}{\xi^2+\omega_{
\uparrow\downarrow}^2} \\\nonumber
&\qquad\times\vect{m}_{\downarrow \uparrow} \cdot  \vect{\nabla} 
\times\ten{G}^{(1)}(\vect{r}_\mathrm{N},\vect{r}_\mathrm{N},\omega) 
\times \overleftarrow{\vect{\nabla'}}\cdot 
\vect{m}_{\uparrow\downarrow}\,,\\\label{ud}
&U_{\uparrow\downarrow}=-U_{\downarrow\uparrow}\\\nonumber
&\quad+\mu_0\vect{m}_{\uparrow \downarrow}\cdot \vect{\nabla} \times 
\mathrm{Re}\ten{G}^{(1)}(\vect{r}_\mathrm{N},\vect{r}_\mathrm{N},
\omega_{\uparrow \downarrow}) \times \overleftarrow{\vect{\nabla'}} 
\cdot \vect{m}_{\downarrow \uparrow},
\end{align}
where the $\vect{m}_{ij}=\left\langle i \right| \hat{\vect{m}}\left| j
\right\rangle$ are the magnetic dipole-matrix elements and
$\ten{G}^{(1)}$ is the scattering part of the Green's tensor. \new{We have used the decomposition \mbox{$\ten{G}= \ten{G}^{(0)}+\ten{G}^{(1)}$} where the translationally-invariant free-space Green's tensor $\ten{G}^{(0)}$ contributes a position-independent energy shift that can be discarded when considering the Casimir--Polder force \cite{Buhmann12}.} The
former can be found by means of rotation operators \cite{sakurai} and
are given by
\begin{align}
&\vect{m}_{\uparrow\uparrow} = -\vect{m}_{\downarrow
\downarrow}=\dfrac{\hbar\gamma_\mathrm{N}}{2}(\sin\vartheta,0,
\cos\vartheta)\,, \\
&\vect{m}_{\uparrow\downarrow} = \vect{m}_{\downarrow
\uparrow}^*=\dfrac{\hbar\gamma_\mathrm{N}}{2}(\cos\vartheta,-i,
-\sin\vartheta)\,.
\end{align}

%%%%%%%%%%%%%%%%%%%%%%%%%%%%%%%%%%%%%%%%%%%%%

In order to calculate the potential further, one has to employ the
Green's tensor corresponding to the setup's geometry. In the case of
the half space, it reads \cite{Buhmann12}
\begin{multline}
\ten{G}^{(1)}(\vect{r},\vect{r}',\omega)=\dfrac{i}{8\pi^2}\\ \cdot
\int\dfrac{\mathrm{d}^2\vect{k}^\parallel}{k^\perp}\sum_{\sigma=s,p}
r_\sigma\vect{e}_{\sigma+}\vect{e}_{\sigma-}e^{i[\vect{k}
^\parallel\cdot(\vect{r}-\vect{r}')+k^\perp(z+z')]}\,,
\end{multline}
where $\vect{k}^\parallel$ and
$k^\perp=\sqrt{(w^2)/(c^2)-{\vect{k^\parallel}}^2}$ are the components
of the wave vector $\vect{k}$ which are parallel and perpendicular to
the interface. The incident (-) and reflected (+) plane waves as 
represented by the polarization unit vectors $\vect{e}_{\sigma\pm}$ 
are polarized parallel ($\sigma=s$) or perpendicular ($\sigma=p$)
to the interface and are reflected according to the respective
Fresnel reflection coefficients $r_\sigma$.

%%%%%%%%%%%%%%%%%%%%%%%%%%%%%%%%%%%%%%%%%%%%%

\subsection{Perfect conductor}
For a perfectly conducting plate with
$r_s=-r_p=-1$, the potential components (\ref{dduu}) and (\ref{du})
simplify to
\begin{align}\label{udpc}
&U_{\downarrow
\uparrow}=\frac{\hbar^2\gamma_\mathrm{N}^2\mu_0}{256\pi^2z^3}
\!\int\limits_{0}^{\infty}\!\frac{\mathrm{d}\xi\,\omega_{
\uparrow\downarrow}}{\omega_{\uparrow\downarrow}^2+\xi^2}[f(\tfrac{\xi
z}{c})+\cos2\vartheta g(\tfrac{\xi z}{c})]e^{-2\xi z/c},\\
&U_{\downarrow\downarrow}=\frac{\hbar^2
\gamma_\mathrm{N}^2\mu_0}{256\pi z^3}\left(1+\cos^2\vartheta\right)\,.
\end{align} 
with $f(x)=5+10x+12x^2$ and $g(x)=-1-2x+4x^2$. The mixed potential
(\ref{udpc}) exhibits two different asymptotes in the retarded,
$(\omega_{\uparrow\downarrow}z)/c\gg1$, and the
nonretarded regimes, $(\omega_{\uparrow\downarrow}z)/c\ll1$.
They read
\begin{align}
&U^\mathrm{ret}_{\downarrow
\uparrow}=\dfrac{\hbar^2\gamma_\mathrm{N}^2\mu_0
c}{32\pi^2\omega_{\uparrow\downarrow}z^4}\,,\\
&U^\mathrm{nret}_{\downarrow \uparrow}=
\dfrac{\hbar^2\gamma_\mathrm{N}^2\mu_0}{512\pi z^3}
\left(5-\cos2\vartheta\right)\,.
\end{align}

\new{The potential near a perfect conductor in the nonretarded regime is independent of the magnitude and direction of the external magnetic field and hence also of the neutron spin. It reads}
\begin{equation}
\new{U_\mathrm{N}}=\dfrac{\hbar^2\gamma_\mathrm{N}^2\mu_0}{64\pi z^3}\equiv\frac{C_3^\text{N}}{z^3}\,.
\end{equation}
\new{This result formally agrees with the repulsive potential of a paramagnetic atom in from of a perfectly conducting plate \cite{Buhmann12}, 
\begin{equation}
U_\mathrm{m}=\frac{\mu_0 }{48 \pi z^3} \left\langle \hat{\vect{m}}^2 \right\rangle,
\end{equation}
}see also the results of Babiker and Barton for an 
arbitrary spin particle \cite{Babiker72}.
For comparison, the potential of an atom $A$ in front of a perfectly
conducting plate reads \cite{0022}
\begin{equation}
U_\mathrm{A}=
-\frac{\langle\hat{\vect{d}}^2\rangle}{48\pi\varepsilon_0
z^3}\equiv-\frac{C_3^\mathrm{A}}{z^3}\,,
\end{equation}
where \new{$C_3^\mathrm{A}=\langle
\hat{\vect{d}}^2\rangle/(48\pi\varepsilon_0)$} is the atomic van
der Waals coefficient and $\hat{\vect{d}}$ its electric dipole
moment. In an order-of-magnitude estimate, one has $\langle
\hat{\vect{d}}^2 \rangle= e^2a_B^2$ where
$a_B=(4\pi\varepsilon_0\hbar^2)/(m_e e^2)$ is the Bohr radius and
$m_e$ is the mass of the electron. Employing
$\gamma=(g_\mathrm{N}e)/(2m_\mathrm{N})$, with
$g_\mathrm{N}=-3.8$ \cite{stocker1994taschenbuch} being the $g$-factor
of the neutron and $m_\mathrm{N}$ being its mass, we find
\begin{equation}
C_3^\mathrm{N}=\dfrac{3}{16}g_\mathrm{N}^2\left(\dfrac{m_e}{m_\mathrm{
N}} \right)^2\alpha^2 C_3^\mathrm{A}\approx
1.7\cdot10^{-10}C_3^\mathrm{A}\,,
\end{equation} 
with $\alpha=e^2/(4\pi c\varepsilon_0 \hbar)\approx 1/137$ being the
fine-structure constant. The Casimir--Polder potential of a neutron
in front of a perfectly conducting plate is hence ten orders of
magnitude smaller than the corresponding potential of a typical atom,
which is due to the smallness of the fine structure constant
(accounting for roughly four orders) and the small
electron-to-neutron mass ratio (accounting for the remaining six
orders).

%%%%%%%%%%%%%%%%%%%%%%%%%%%%%%%%%%%%%%%%%%%%%

\subsection{Metals and dielectrics}
\begin{figure*}
\centering
\includegraphics[width=0.6\textwidth]{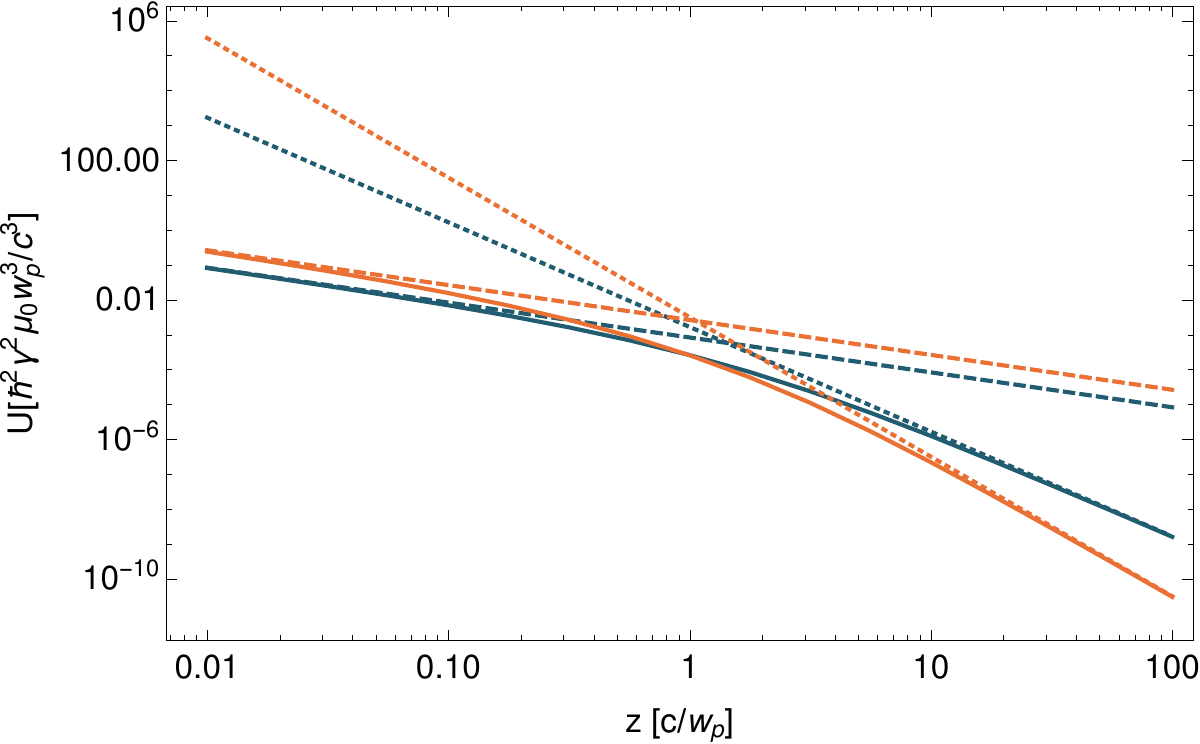}
\caption{Constituents of the Casimir--Polder potential of a
ground-state (spin-down) neutron in front of a surface described by a
plasma model with $\omega_{\uparrow\downarrow}=\omega_P$:
$U_{\downarrow\uparrow}$ (orange) and $U_{\downarrow\downarrow}$
(blue), exact potential (solid), retarded asymptote (dotted) and
nonretarded asymptote (dashed).}
\label{fig:1}
\end{figure*}

To be more realistic, we describe the electric response of the
plate by the plasma model,
\mbox{$\varepsilon=1-\omega_\mathrm{P}^2/\omega^2$}, the Drude model
as given by
\mbox{$\varepsilon=1-\omega_\mathrm{P}^2/[
\omega(\omega+\mi\gamma\omega) ] $}
and a single-resonance Drude--Lorentz model,
\mbox{$\varepsilon=1-\omega_\mathrm{P}^2/(\omega^2-\omega_\mathrm{T}
^2)$}. \new{Averaging over the direction of the external magnetic field,} we find a repulsive ground-state
potential for each of these models. As an illustration, we show its constituents and the
respective retarded and nonretarded limits for the plasma
model in Fig.~\ref{fig:1}.

%%%%%%%%%%%%%%%%%%%%%%%%%%%%%%%%%%%%%%%%%%%%%

\section{Discussion}
In an experiment, applied magnetic fields are typically
$B_\mathrm{ext}\lesssim 5\mathrm{T}$, such that the critical distance
\mbox{$z_\mathrm{nret}=c/\omega_{\uparrow\downarrow}
=c/(\gamma_\mathrm{N}B_\mathrm{ext})\gtrsim 0.32\,\mathrm{m}$} for the
nonretarded limit is much larger than typical distances in
experiments which vary from $\mathrm{nm}$ to $\mu\mathrm{m}$.
We have $z\ll z_\mathrm{nret}$, such that we find ourselves in the
nonretarded limit. Since each model exhibits its own characteristic 
frequencies, one can identify different asymptotes for each model 
where the potential can be described by power laws. The
asymptotes together with the corresponding ranges of validity are 
summarised in Tab.~\ref{tab:1}. The dispersive parts of these 
asymptotes with their analogues for magnetic atoms which have 
previously been reported for metals \cite{Henkel05, Haakh09} and 
dielectrics \cite{0831} with two exceptions: our 
\mbox{$z\ll z_\mathrm{critical}$} asymptote for a Drude metal differs 
from that given in Refs.~\cite{Henkel05, Haakh09}, because we do 
not employ a high-frequency cut-off, whereas the 
\mbox{$z\gg z_\mathrm{critical}$} regime for dielectrics is very 
specific to neutron interactions and has hence not been considered 
for 
atoms.

\begin{table*}[t]
\centering
\begin{tabular}{c|c|c|c}
Model          &$\quad U_\mathrm{N}(z\ll z_\mathrm{critical})$	& 
$\quad U_\mathrm{N}(z\gg z_\mathrm{critical})$ & $z_\mathrm{critical}$
 \\\hline
I    & 
\multicolumn{2}{c|}{\begin{minipage}{25ex}\centering\vspace*{1ex}
$\displaystyle
\frac{\hbar^2\gamma_\mathrm{N}^2\mu_0}{64\pi z^3}$
\vspace*{1ex}\end{minipage}} &$\displaystyle
-$\\\hline
II       & 
\begin{minipage}{29ex}\centering\vspace*{1ex}$\displaystyle
\frac{\hbar^2\gamma_\mathrm{N}^2\mu_0\omega_p^2}{128\pi c^2 z}$
\vspace*{1ex}\end{minipage}&  
\begin{minipage}{ 25ex}\centering\vspace*{1ex}
$U_\mathrm{N}(\mbox{Perf.\ Cond.})$ \vspace*{1ex}\end{minipage} 
&$\displaystyle
\frac{c}{\omega_\mathrm{P}}$\\\hline
III         &\begin{minipage}{29ex}\centering\vspace*{1ex}
$\displaystyle-\frac{\hbar^2\gamma_\mathrm{N}
^2\mu_0\omega_p^2}{96\pi^2c^2
z}\frac{\gamma_\mathrm{N}
B_\mathrm{ext}}{\gamma}\ln \frac{\gamma_\mathrm{N}
B_\mathrm{ext}}{\gamma}$ \vspace*{1ex}\end{minipage} & 
\begin{minipage}{25ex}\centering\vspace*{1ex}
$\frac{2}{3}U_\mathrm{N}(\mbox{Perf.\ 
Cond.})$\vspace*{1ex}\end{minipage} & $\displaystyle 
\frac{c}{\omega_\mathrm{P}}\sqrt{\frac{\gamma}{\gamma_\mathrm{N} 
B_\mathrm{ext}}}$\\\hline
IV &\begin{minipage}{29ex}\centering\vspace*{1ex} 
$\displaystyle
\frac{\hbar^2\gamma_\mathrm{N}^2\mu_0\omega_\mathrm{P}^2}{192\pi
c^2}\frac{\omega_\mathrm{T}+\omega_\mathrm{L}}{\omega_\mathrm{T}
\omega_\mathrm{L}}\dfrac{1}{z} \gamma_\mathrm{N} B_\mathrm{ext}$
\vspace*{1ex}\end{minipage} & 
\begin{minipage}{25ex}\centering\vspace*{1ex}
$\displaystyle \frac{\hbar^2\gamma_\mathrm{N}
^2\mu_0\gamma_\mathrm{N}B_\mathrm{ext}}{576\pi^2c
z^2}f(\omega_\mathrm{T},\omega_\mathrm{L})$\vspace*{1ex}\end{minipage} 
&$\displaystyle
\frac{c}{\omega_\mathrm{T}}$ \end{tabular}
\caption{Nonretarded ground-state potential $U_\mathrm{N}$ for 
different models (I: Perfect conductor, II: Plasma, III: Drude, IV: Drude--Lorentz) within leading order in the applied magnetic field, 
valid for $z\ll c/(\gamma_\mathrm{N}B_\mathrm{ext})$.  
$\omega_\mathrm{L}=\sqrt{\omega_\mathrm{T}^2+\omega_\mathrm{P}^2/2}$ 
and 
$f(\omega_\mathrm{L},\omega_\mathrm{T})=(\omega_\mathrm{L}
^4\omega_\mathrm{T}^2+5\omega_\mathrm{L}^6-3\omega_\mathrm{L}
^2\omega_\mathrm{T}^4-\omega_\mathrm{T}^6)/(\omega_\mathrm{L}
^4\omega_\mathrm{T}^2)$.}
\label{tab:1}
\end{table*}

\normalsize

The asymptotes show that for the perfect conductor and the plasma 
model, the potential persists even of vanishing external magnetic 
field. For the two other models instead, the potential vanishes in the 
limit of vanishing external magnetic field. We note that the different 
models lead to a variety of asymptotic power laws for the distance 
dependence. Interestingly, it is seen that the perfect-conductor limit 
does not commute with the nonretarded limit for the plasma model, in 
contrast to the case of the Casimir--Polder potential of an electrically polarisable atom
\cite{0296}. In addition, there are marked differences between the 
plasma and Drude models, making this interaction a new and sensitive 
test case for the Drude--plasma debate in Casimir physics 
\cite{Brevik06,Bordag09}. Ultimately, the strong model-dependence of 
the neutron Casimir--Polder potential stems from its mixed 
electric--magnetic interaction in a short-distance regime, which is 
analogous to the case of the anomalous magnetic moment of the electron 
\cite{Bennett13,Bennet18}.

\begin{figure*}[t]
\centering
\includegraphics[width=0.6\textwidth]{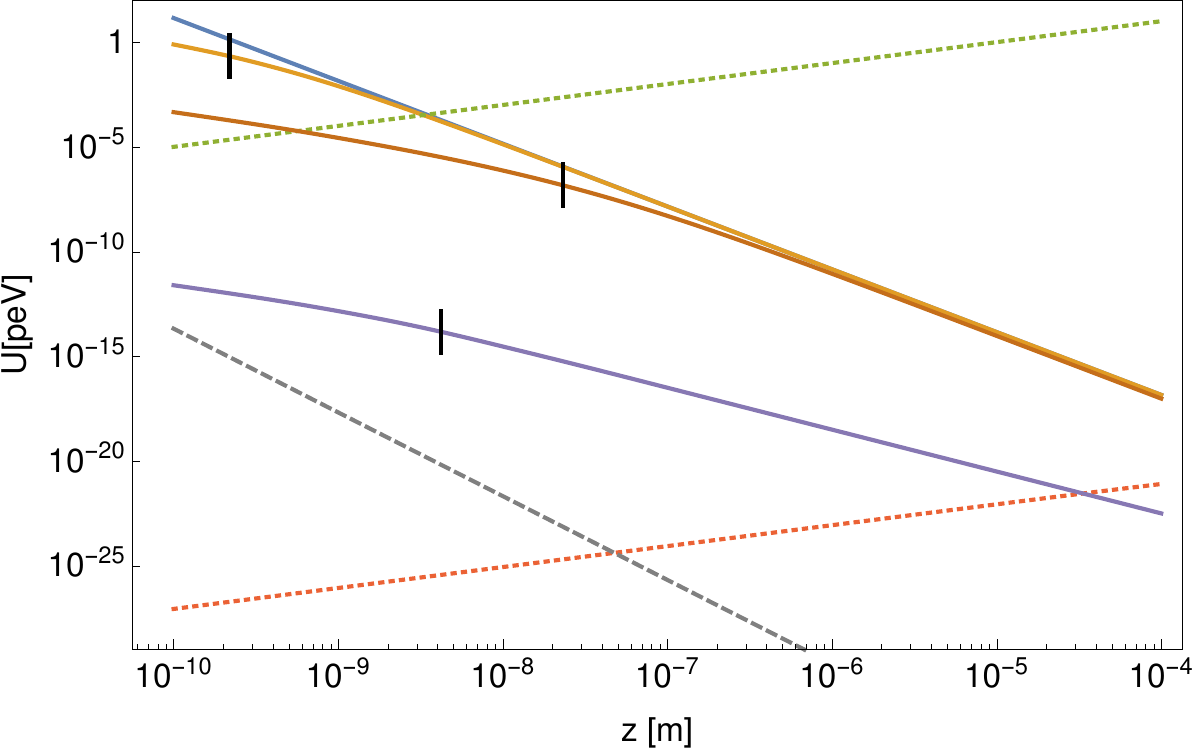}
\caption{Ground-state Casimir--Polder potential of a neutron in front
of a surface with applied field \mbox{$B_\mathrm{ext}=2\,\mathrm{T}$}
(solid lines) compared to gravitational potentials (dashed) \new{and the absolute value of the attractive potential due to the neutron's static electric and magnetic polarisability (dashed)}: Perfect
conductor (blue), plasma model for gold (orange,
\mbox{$\omega_\mathrm{P}=1.37\cdot10^{16}\,\mathrm{rad}/\mathrm{s}$}
\cite{palik1998handbook}), Drude model for gold (red,
\mbox{$\gamma=4.10\cdot10^{12}\,\mathrm{rad}/\mathrm{s}$}
\cite{palik1998handbook}), Drude--Lorentz model for silicon (purple,
\mbox{$\omega_\mathrm{P}=2.3\cdot10^{16}\,\mathrm{rad}/\mathrm{s}$},
\mbox{$\omega_\mathrm{T}=7.1\cdot10^{16}\,\mathrm{rad}/\mathrm{s}$}
\cite{Pirozenko08}),
gravitational potentials of Earth (green) and of a silicon sphere 
(red). The black vertical lines represent the critical distances for 
each model.}
\label{fig:2}
\end{figure*}

Finally, let us discuss whether the potentials predicted for the 
different models are in principle observable in an experiment. To
that end, we compare them with the gravitational potential exerted on
the neutron by the same plate and by Earth, respectively\new{, as well as the potential arising from the static electric and magnetic polarisabilities of the neutron}. In 
Fig.~\ref{fig:2}, the neutron Casimir--Polder potential at accessible 
distance regimes is shown for the different models alongside \new{the}
gravitational \new{and static} potentials. We see for instance that both the plasma and 
the Drude model approach the perfect conductor at different critical 
distances but always are smaller than the latter. The Drude--Lorentz 
model gives rise to the smallest potential at all distances. The 
magnitude of the Casimir-Polder potential is highly model-dependent. 

Furthermore, while the Casimir--Polder potential is generally weaker than the gravitational potential of Earth 
except for the perfect-conductor case, it is for all models stronger 
than the gravitational potential of the plate itself. It should hence 
be taken into account when performing short-range gravity experiments 
with neutrons. To estimate the gravitational field of a surface in a 
typical neutron interferometry experiment \cite{Rauch02}, we have for 
simplicity used a silicon sphere (density 
$\rho=2.33\,\mathrm{g}/\mathrm{cm}^3$) of radius $r=11.3\,\mathrm{mm}$ 
whose mass is comparable to a plate in a perfect crystal 
interferometer. For another state-of-the-art neutron setup probing 
Earth's gravitation, see e.g. reference~\cite{Kulin15}. 

\new{Static electric and magnetic polarizabilities of the neutron result in the potential \cite{0095}
\begin{equation}
 U_\mathrm{static}=-\frac{3\hbar c (\alpha - \beta)}{8 \pi z^4},
\end{equation}
where $\alpha$ and $\beta$ are the neutron's static electric and magnetic polarisabilities, respectively ($\alpha=(11.8\pm1.1)\cdot 10^{-4}\mathrm{fm}^3$, $\beta=(3.7\pm1.2)\cdot 10^{-4}\mathrm{fm}^3$ \cite{Zyla20}). The corresponding potential is negative and thus attractive. Its absolute value is shown as a dashed line in Fig.~\ref{fig:2}. We see that the potential due to static polarisabilities is smaller than the repulsive potential derived in our work, by several orders of magnitude.}

By contrast, in gravitational resonance spectroscopy experiments, ultra-cold neutrons form gravitationally bound states above a planar surface which is totally reflecting for neutrons. Currently, transitions between such 
states can be resolved with $\Delta E \leq 2 \cdot 10^{-3}\,$~peV 
\cite{Cronenberg2018}, while near-term experimental setups are expected to achieve $\Delta E \sim  10^{-5}\,$~peV \cite{Sedmik2020}. The wave functions of the lowest states 
probe the space region up to $60\,\mu\mathrm{m}$ above the mirror, 
with energies in the $\mathrm{peV}$ range, see Appendix A for details. The first-order 
perturbative energy shift of the gravitational state $\psi_n(z)$ due 
to the Casimir--Polder potential of a \new{mirror described by the plasma model} reads
\begin{eqnarray}
\delta E_{\mathrm{CP}}(n) 
&=& \int_{z=0}^\infty\dif z\, |\psi_n(z)|^2
\new{ U^\mathrm{plasma}_\mathrm{N}(z)}.
%&=&\frac{\hbar^2\gamma_\mathrm{N}^2\mu_0}{64 \pi}\int_{z=0}^{\infty}\dif z\, 
%\frac{|\psi_n(z)|^2}{z^3}\,. 
\label{eq:maintrans2}
\end{eqnarray}
\new{Note that the integral in equation (\ref{eq:maintrans2}) converges due to the behaviour of the neutron's wavefunction for small distances. This is not the case when using the perfect conductor model which has to be regularized by including the lattice constant of the half-space medium as a cut-off for small distances, beyond which the continuoum description of the perfect-conductor model breaks down.} 
For currently realized setups \cite{Cronenberg2018,Cronenberg2015}, the 
Casimir--Polder interaction predicted by the \new{plasma} model 
would lead to an energy shift of $\Delta E_{\mathrm{CP}}(n,m)=\delta 
E_{\mathrm{CP}}(n)-\delta E_{\mathrm{CP}}(m)=\new{1.7}\cdot10^{-14}\, \mathrm{peV}$ for the transition $(n,m)=(1,5)$, which is below the detection threshold. This is 
true even for a planned improved setup involving Ramsey-based gravity 
resonance spectroscopy \cite{Abele2010} with an estimated sensitivity of 
\mbox{$\Delta E = 5 \cdot 10^{-9}\,\mathrm{peV}$} which is 
statistically limited. 
The influence of the Casimir--Polder potential can be increased by adding an additional mirror on top, separated from the lower mirror by a distance of a few tens of microns.
The two mirrors effectively squeeze the neutron's wavefunction, leading to an increase of the energy shifts in the transitions by orders of magnitude. For instance, the energy shift of the transition $(1,5)$ is increased by a factor of $200$ (taking a typical mirror spacing of $15\,\mu\mathrm{m}$), see appendix A. While this energy shift still remains under the current detection threshold, future experiments with a sensitivity 
below the expected energy shift could distinguish between these models. In this case, contributions from the Casimir--Polder potential have to be taken into account. Furthermore, setups whose bound states have a larger 
neutron density close to the surface would lead to enhanced 
Casimir--Polder shifts that are cubically enhanced with the inverse 
distance.

%%%%%%%%%%%%%%%%%%%%%%%%%%%%%%%%%%%%%%%%%%%%%

\section{Conclusions}
We have shown that a single neutron under the influence of a constant
magnetic field will be subject to a repulsive Casimir--Polder-type dispersion
interaction with a metal or dielectric plate. This is a rare example of
Casimir repulsion for a magnetisable object interacting with a
polarisable one where the repulsion is not dominated by an attractive
electric--electric force. We have found that the force is
nonretarded for experimentally accessible regimes and that it is
very sensitive to the electric response of the surface. It may hence
provide a testing ground for the Drude-plasma debate. In addition, while typically smaller than Earth's gravitational potential by orders of magnitude, it can become comparable to the gravitational interaction of the same surface. Furthermore, we find that while in current neutron experiments, the contribution of Casimir--Polder interactions is not visible, in future high-precision gravitational experiments on short length-scales, a possible influence has to be taken into account.

\section*{Acknowledgements}
We are grateful for discussions with G.~Barton, A.~Buchleitner, R.~Decca, 
F.~Intravia, H.~Lemmel and H.~Rauch. This work was supported by the 
German Research Foundation (DFG, grants BU 1803/3-1 and GRK 2079/1) and the Austrian Science 
Fund (FWF, Doctoral program Particles \& Interactions, project no. 
W1252).

\appendix

\section{ Energy shifts in Gravitational Resonance Spectroscopy experiments}

\begin{figure*}[t]
\centering
\includegraphics[width=0.8\textwidth]{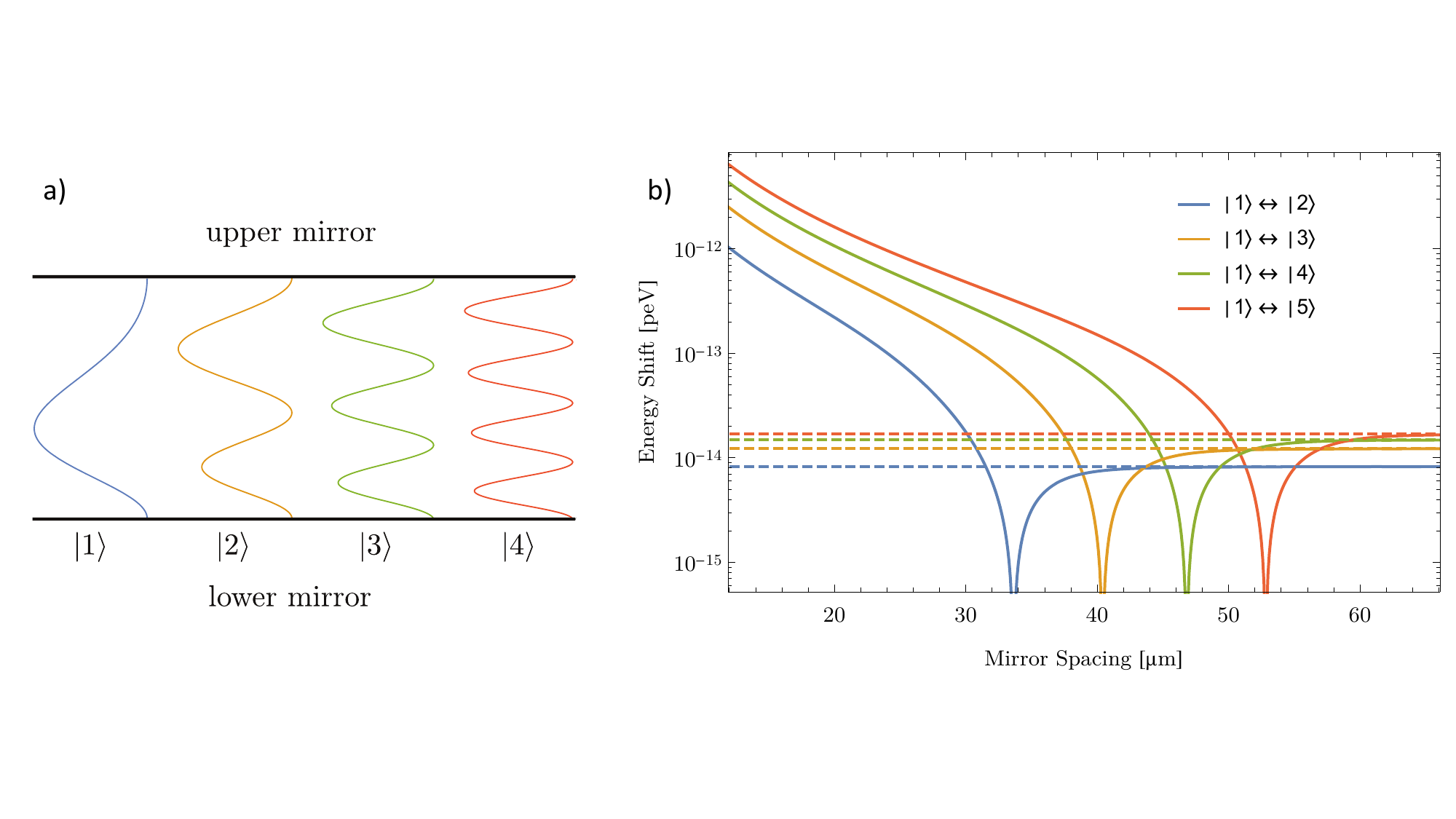}
\caption{(a) Sketch of the lowest energy bound states of the neutrons in the double mirror configuration. (b) Corrections $|\Delta E_{\mathrm{CP}}(n,m)|$ of the transition energies due to the Casimir--Polder interaction of the neutron in the double mirror configuration as a function of the mirror spacing. Dashed lines indicate the correction of the transition energies for a single-mirror setup.  The mirrors are modeled with the plasma model (\mbox{$\omega_\mathrm{P}=1.37\cdot10^{16}\,\mathrm{rad}/\mathrm{s}$}). }
\label{fig:3}
\end{figure*}

Here we discuss in more detail the influence of the Casimir--Polder potential in Gravitational Resonance Spectroscopy experiments. As discussed above, different gravitational bound states of the neutron experience different energy shifts due to the Casimir--Polder interaction of the neutron with nearby mirrors. For a neutron localised above a totally reflecting mirror, the $n$-th bound state is described by the wavefunction \cite{Westphal2007}
\begin{equation}
 \psi_n(z) =c_A(n)\operatorname{Ai}(\hat z-E(n)), 
\end{equation}
where $\operatorname{Ai}$ denotes the Airy function of the first kind, $c_A(n)$ is a normalization constant, $\hat z$ is the rescaled distance to the mirror and $ E(n)$ is the rescaled eigenenergy of the bound state. In the case of a double mirror configuration with mirror spacing $l$, the bound wavefunctions are given by \cite{Westphal2007}
\begin{equation}
 \psi_n(z) =c_A(n,l)\operatorname{Ai}(\hat z-E(n,l))+c_B(n,l)\operatorname{Bi}(\hat z-E(n,l)), 
\end{equation}
where $\operatorname{Bi}$ is the Airy function of second kind and the rescaled eigenenergies $E(n,l)$ as well as the coefficients $c_A(n,l)$ and $c_B(n,l)$ now depend on the mirror spacing $l$. A sketch of the wavefunctions for the lowest energy bound states in the double mirror configuration is shown in Fig.~\ref{fig:3}(a). Note that a penetration of the wavefunctions into the mirror is neglected.

Using equation (\ref{eq:maintrans2}), we can now compute first-order energy corrections for transitions between the different bound states. Modeling the reflecting mirrors with the plasma model, we show different transition energies as a function of the mirror spacing $l$ in Fig.~\ref{fig:3}(b). We see that by varying the mirror spacing, one can manipulate the influence of the Casimir--Polder effect on the different transitions. In particular, for small mirror spacings, the effect is increased by several orders of magnitude with respect to the correction in case of a single-mirror setup (dashed lines).


\begin{thebibliography}{99}
%
\bibitem{0373}
H.~B.~G.\ Casimir, Proc.\ K.\ Ned.\ Akad.\ Wet.\ \textbf{51},
793 (1948).

\bibitem{0057}
E.~M.\ Lifshitz, Sov.\ Phys.\ JETP \textbf{2}, 73 (1956).

\bibitem{0487}
P.~W.\ Milonni, \textit{The Quantum Vacuum} (Academic Press, New York,
1994).

\bibitem{0030}
H.~B.~G.\ Casimir and D.~Polder, Phys.\ Rev.\ \textbf{73},
360 (1948).

\bibitem{0122}
T.~H.\ Boyer, Phys.\ Rev.\ A \textbf{9}, 2078 (1974).

\bibitem{0123}
V.~Hushwater, Am.\ J.\ Phys.\ \textbf{65}, 381 (1997).

\bibitem{0089}
G.~Feinberg and J.~Sucher, J.\ Chem.\ Phys.\ \textbf{48},
3333 (1968).

\bibitem{0838}
A.~Salam, Int.\ J.\ Quantum Chem.\ \textbf{78}, 437 (2000).

\bibitem{0134}
C.~Henkel and K.\ Joulain, Europhys.\ Lett.\ \textbf{72},
929 (2005).

\bibitem{0126}
O.~Kenneth, I.~Klich, A.~Mann and M.~Revzen, Phys.\ Rev.\
Lett.\ \textbf{89}, 033001 (2002).

\bibitem{0095}
T.~H.\ Boyer, Phys.\ Rev.\ \textbf{180}, 19 (1969).

\bibitem{0330}
Y.~Tikochinsky and L.~Spruch, Phys.\ Rev. A\ \textbf{48},
4236 (1993).

\bibitem{Henkel05}
C.~Henkel, B.~Power and F.~Sols, J.\ Phys.\ Conf. Ser. \textbf{19}, 34
(2005).

\bibitem{0019}
S.~Y.\ Buhmann, D.-G.\ Welsch and T.~Kampf, Phys.\ Rev.\ A
\textbf{72}, 032112 (2005).

\bibitem{0831}
H.~Safari, D.-G.\ Welsch, S.~Y.\ Buhmann and S.~Scheel,
Phys.\ Rev. A\ \textbf{78}, 062901 (2008).

\bibitem{Haakh09}
H.~Haakh, F.~Intravaia, C.~Henkel, S.~Spagnolo, R.~Passante, B.~Power 
and F.~Sols, Phys.\ Rev.\ A, \textbf{80}, 062905 (2009).

\bibitem{Bimonte09}
G.~Bimonte, G.~L.\ Klimchitskaya and V.~M.\ Mostepanenko
Phys.\ Rev.\ A \textbf{79}, 042906 (2009).

\bibitem{0022}
J.~E.\ Lennard-Jones, Trans.\ Faraday Soc.\ \textbf{28},
333 (1932).

\bibitem{0096}
C.~Farina, F.~C.\ Santos and A.~C.\ Tort, J.\ Phys.\ A:
Math.\ Gen.\ \textbf{35}, 2477 (2002).

\bibitem{0121}
C.~Farina, F.~C.\ Santos and A.~C.\ Tort, Am.\ J.\ Phys.\
\textbf{70}, 421 (2002).

\bibitem{0578}
F.~M.\ Serry, D.~Walliser and G.~J.\ Maclay, J.\ Appl.\
Phys.\ \textbf{84}, 2501 (1998).

\bibitem{0961}
J.~N.\ Munday, F.~Capasso and V.~A.\ Parsegian, Nature
\textbf{457}, 170 (2009).

\bibitem{0771}
J.~M.\ Obrecht, R.~J.\ Wild, M.~Antezza, L.~P.\
Pitaevskii, S.~Stringari and E.~A.\ Cornell, Phys.\ Rev.\ Lett.\
\textbf{98}, 063201 (2007).

\bibitem{0157}
H.~Failache, S.~Saltiel, M.~Fichet, D.~Bloch, M.~Ducloy, Phys.\
Rev.\ Lett.\ \textbf{83}, 5467 (1999).

\bibitem{0953}
F.~S.~S.\ Rosa, D.~A.~R.\ Dalvit and P.~W.\ Milonni, Phys.\
Rev.\ Lett.\ \textbf{100}, 183602 (2008).

\bibitem{0954}
F.~S.~S. Rosa, D.~A.~R. Dalvit and P.~W.\ Milonni, Phys.\
Rev.\ A \textbf{78}, 032117 (2008).

\bibitem{0946}
S.~J.\ Rahi, M.~Kardar and T.~Emig, Phys.\ Rev.\ Lett.\ \textbf{105},
070404 (2010).

\bibitem{Babiker72}
M.~Babiker and G.~Barton, Proc.\ R.\ Soc.\ Lond.\ Ser.\ A 
\textbf{326}, 255 (1972). 

\bibitem{Babb16} 
J.~F.\ Babb and M.~S.\ Hussein, EPJ Web of Conferences \textbf{113}, 
08001 (2016).

\bibitem{Baumann:1988}
J.~Baumann, R.~Gahler, J.~Kalus and W.~Mampe, Phys.\
Rev.\ D \textbf{37}, 3107 (1988).

%\bibitem{Pendlebury2015} 
%J.~M.\ Pendlebury \textit{et al}, Phys.\
%Rev.\ D \textbf{92}, 092003 (2015).

\bibitem{Abel2020}
C.~Abel \textit{et al}, Phys.\ Rev.\ Lett. \textbf{124}, 081803 (2020).

\bibitem{Lins:2015}
I.~Altarev \textit{et al}, (2015), J.\ Appl.\ Phys. \textbf{117}, 183903 (2015).

\bibitem{Rauch02}
H.~Rauch, H.~Lemmel, M.~Baron and R.~Loidl, Nature \textbf{417},
630 (2002).

\bibitem{Buhmann12}
S.~Y.\ Buhmann, \textit{Dispersion Forces I---Macroscopic Quantum
Electrodynamics and Ground-State Casimir, Casimir--Polder and van der
Waals Forces} (Springer, Heidelberg, 2012).

\bibitem{sakurai}
J.~J.\ Sakurai, \textit{Modern Quantum Mechanics} (Addison--Wesley,
New York, 1994).

\bibitem{stocker1994taschenbuch}
H.~St\"ocker, \textit{Taschenbuch der Physik}, 2nd Ed.
(Harri Deutsch, Frankfurt, 1994).

\bibitem{0296}
M.~Babiker and G.~Barton, J.\ Phys.\ A:\ Math.\ Gen.\
\textbf{9}, 129 (1976).

\bibitem{Brevik06}
I.~Brevik, S.~A.\ Ellingsen, K.~A.\ Milton, New J.\ Phys.\ \textbf{8},
236 (2006).

\bibitem{Bordag09}
M. Bordag, G. L. Klimchitskaya, U. Mohideen, V. M. Mostepanenko,
\textit{Advances in the Casimir effect} (Oxford University Press,
Oxford, 2009) and references therein.

\bibitem{Bennett13}
R.~Bennett and C.~Eberlein, Phys. Rev. A \textbf{88}, 012107 (2013).

\bibitem{Bennet18}
R.~Bennett, S.~Y.\ Buhmann and C.~Eberlein, Phys. Rev. A \textbf{98}, 022515 (2018).

\bibitem{palik1998handbook}
Palik, E. D. (ed.), \textit{Handbook of Optical Constants of Solids
III} (Academic Press, New York, 1991).

\bibitem{Pirozenko08}
I.~Pirozhenko and A.~Lambrecht, Phys. Rev. A \textbf{77}, 013811
(2008).

\bibitem{Kulin15}
G.~V.\ Kulin, A.~I.\ Frank, S.~V.\ Goryunov, D.~V.\ Kustov, 
P.~Geltenbort, M.~Jentschel, A.~N.\ Strepetov and V.~A.\ Bushuev,
Nucl.\ Instrum.\ Meth.\ Phys.\ Res.\ A \textbf{792}, 38 (2015).


\bibitem{Zyla20}
P.~A.\ Zyla \textit{et al}, Prog. Theor. Exp.
Phys. \textbf{2020}, 083C01 (2020).

\bibitem{Cronenberg2018}
G.~Cronenberg \textit{et al}, Nat. Phys. \textbf{14} 1022 (2018).

\bibitem{Sedmik2020}
R.~Sedmik \textit{et al}, J. Synch. Investig. \textbf{14}, 195 (2020).

\bibitem{Cronenberg2015} 
G.~Cronenberg, H.~Filter, M.~Thalhammer, T.~Jenke, H.~Abele and P.~Geltenbort, PoS (EPS-HEP2015) 408 (2015).

\bibitem{Abele2010}
H.~Abele, T.~Jenke, H.~Leeb and J.~Schmiedmayer, Phys.\ Rev.\ D
\textbf{81}, 065019 (2010).

\bibitem{Westphal2007}
A.~Westphal, H.~Abele, S.~Bae{\ss}ler, V.~V.\ Nesvizhevsky, K.~V.\ Protasov and A.~Y.\ Voronin, Eur. Phys. J. \textbf{C51},
367 (2007).
%
\end{thebibliography}
\end{document}